\documentclass[twocolumn]{aastex63}  
\usepackage{amssymb,amsmath,bm}
\usepackage{graphicx}
\usepackage{natbib}
\usepackage{color}
\usepackage{algorithm}
\usepackage[noend]{algpseudocode}
\usepackage{tikz}
\usetikzlibrary{quantikz}
\usepackage{savesym}
\savesymbol{splitbox}
\usepackage{adjustbox}
\restoresymbol{TXF}{splitbox}
\usepackage{hyperref}

\newcommand{\dg}{\dagger}

\received{xxx xx, 2021}
\revised{xxx xx, 2021}
\accepted{xxx xx, 2021}
\submitjournal{ApJ}

\begin{document}

%
%
%
%


\title{Towards Cosmological Simulations of Dark Matter on Quantum Computers}

\correspondingauthor{Philip Mocz}
\email{pmocz@astro.princeton.edu}

\author[0000-0001-6631-2566]{Philip Mocz}
\affiliation{Department of Astrophysical Sciences, Princeton University, 4 Ivy Lane, 
Princeton, NJ 08544, USA}

\author[0000-0002-1127-2111]{Aaron Szasz}
\affiliation{Perimeter Institute for Theoretical Physics, Waterloo, Ontario N2L 2Y5, Canada}
\affiliation{Materials Sciences Division, Lawrence Berkeley National Laboratory, Berkeley, CA 94720, USA}


\begin{abstract}
State-of-the-art cosmological simulations on classical computers are limited by time, energy, and memory usage.  Quantum computers can perform some calculations exponentially faster than classical computers, using exponentially less energy and memory, and may enable extremely large simulations that accurately capture the whole dynamic range of structure in the Universe within statistically representative cosmic volumes.  However, not all computational tasks exhibit a `quantum advantage'. Quantum circuits act linearly on quantum states, so nonlinearities (e.g. self-gravity in cosmological simulations) pose a significant challenge. Here we outline one potential approach to overcome this challenge and solve the (nonlinear) Schr\"odinger-Poisson equations for the evolution of self-gravitating dark matter, based on a hybrid quantum-classical variational algorithm framework \citep{2020PhRvA.101a0301L}.  We demonstrate the method with a proof-of-concept mock quantum simulation, envisioning a future where quantum computers will one day lead simulations of dark matter.
\end{abstract}


\section{Introduction}
\label{sec:intro}

Quantum computers have enormous potential to exponentially speed up classical algorithms by leveraging the quantum mechanical features of \textit{superposition} and \textit{entanglement} \citep{1982IJTP...21..467F}.
The elementary unit of a quantum computer is a \textit{qubit} (a quantum version of the classical bit). Whereas $n$ classical bits can encode $2^n$ integers, $n$ qubits instead describe a vector space where each of these integers corresponds to one basis state. By acting on more general vectors (superpositions), operations can be performed in parallel on all basis states,
with physical costs (energy) required for computation that only scale as $n$ \citep{2008AmJPh..76..657B}.
Quantum computers, as they become a reality (\citealt{sevilla2020forecasting}, e.g. \citealt{2019Natur.574..505A}),
will open new avenues for 
direct simulation of quantum systems \citep{2008PNAS..10518681K}, 
factoring integers \citep{1995quant.ph..8027S},
and solving systems of linear equations \citep{2009PhRvL.103o0502H},  
among other applications \citep{2018arXiv180403719A}.
However, not all computational tasks are amenable to exponential speedup on a quantum computer. Limitations include (1) the unitarity and reversibility of quantum gates, (2) the impossibility of copying an arbitrary unknown quantum state (the \textit{no-cloning} theorem; \citealt{1970FoPh....1...23P}), and (3) that output states are accessible only through measurement, making computational results probabilistic in nature.  
More broadly, computational complexity theory tells us that some problems are simply too hard to solve for any reasonable model of computation.

Whether and how, despite these limitations, \textit{nonlinear} equations can be solved with exponential speedup using unitary (\textit{linear}) quantum gates, remains an open research problem.
Nonlinear products can be computed by preparing multiple copies of a quantum state \citep{2008arXiv0812.4423L}, so one could envision first preparing a state, then determining the evolved state at each timestep via measurement, then preparing further copies to use in the next timestep. But this approach would be far worse than a classical simulation, as determining an arbitrary wavefunction requires exponentially many measurements in the number of qubits and retains statistical errors not present in classical computation. Additionally, preparing an arbitrary state is difficult \citep{aharonov2003adiabatic}. Alternatively, the problem of measuring and preparing arbitrary states could be avoided by carrying out many copies of the time evolution from the initial state in parallel, but this would require exponentially many copies in the number of timesteps.  Thus something beyond these naive approaches is needed.
Despite recent progress on solving nonlinear equations by direct simulation \citep{liu2020efficient, lloyd2020quantum}, we focus instead on the use of variational quantum computing (VQC; \citealt{cerezo2020variational}). 

VQC finds the solution to a nonlinear equation using the key concept that the problem can be written as a function of variational parameters, and optimal parameters can be found by minimizing a cost function.  The generally expensive-to-compute cost function is calculated on a quantum computer, while the parameter optimization is done classically; VQC is consequently a hybrid quantum-classical approach.
A VQC method for solving nonlinear differential equations was outlined and demonstrated in \cite{2020PhRvA.101a0301L}
with 2 qubits on the Noisy Intermediate-Scale Quantum (NISQ) IBM-Q device and with a 13 qubit mock simulation using matrix product states \citep{Schollwock2011}. 
The VQC framework has been considered for simulating nonlinear systems in quantum chemistry \citep{2016NatSR...632940K}, Navier-Stokes fluids \citep{griffininvestigation}, and plasma physics \citep{2020arXiv200514369D}.

Nonlinear differential equations underlie simulations of cosmic structure formation, which would benefit from larger simulations. Galaxies form in the self-gravitating (nonlinear) potential wells of dark matter halos.  A statistically representative volume of the Universe requires Gigaparsec box sizes, while galaxies are kiloparsec-scale: $10^6\times$ smaller in linear dimension \citep{2017ComAC...4....2P}. 
Halos span from massive $10^{14}~M_\odot$ structures to potentially Earth-mass subhalos: 19 orders of magnitude \citep{2009NJPh...11j5027B}. 
The small-scale distribution of dark matter may highlight interesting additional physics in the nature of dark matter (e.g. self-interactions) but is inextricably linked to larger scales \citep{2017ARA&A..55..343B}. 
This huge dynamic range is far from achievable by today's classical supercomputers \citep{2014MNRAS.445..175G,2018ApJS..236...43G}, hence technological innovation that allows for larger simulations is essential.

This manuscript's aim is to begin bridging emerging quantum computing technology and applications to cosmological simulations.
We first describe the physical system of interest (\S~\ref{sec:system}) and a classical spectral solver (\S~\ref{sec:classical}).
We present a high-level overview of a VQC algorithm (\S~\ref{sec:quantum}) that in principle would exponentially speed up cosmological simulations of dark matter, along with a proof-of-concept mock simulation (\S~\ref{sec:mock}). Finally, we discuss the method's advantages and limitations and offer some perspective for the near future (\S~\ref{sec:conc}).

\section{Self-gravitating dark matter in an expanding universe}
\label{sec:system}

Consider the evolution of a (non-relativistic) dark matter scalar field in an expanding universe.
In dimensionless form, the evolution of the complex wavefunction $\psi$ describing the dark matter as a function of time $t$ is given by the Schr\"odinger-Poisson equations:
\begin{equation}
\label{eqn:S}
i\frac{\partial}{\partial t}\psi  = -\frac{\nabla^2}{2}\psi + aV\psi
\end{equation}
\begin{equation}
\label{eqn:P}
\nabla^2 V = |\psi|^2-1
\end{equation}
where $a\equiv a(t)$ is the prescribed cosmic scale factor that encodes underlying cosmological expansion.  $\psi$ is normalized over the volume $\mathcal{V}$ such that $\frac{1}{\mathcal{V}}\int|\psi|^2 \,d\mathcal{V}=1$.

The equations are applicable to a range of dark matter models, including ultra-light boson dark matter (`fuzzy dark matter'; \citealt{2000PhRvL..85.1158H,2014NatPh..10..496S,2017PhRvD..95d3541H,2017MNRAS.471.4559M}) as well as standard collisionless cold dark matter (CDM) in the limit that the particle mass tends to infinity.
In the large mass limit, the Schr\"odinger-Poisson equations approximate the Vlasov-Poisson equations \citep{1993ApJ...416L..71W,2017PhRvD..96l3532K,2018PhRvD..97h3519M}, which are the more commonly solved equations in CDM simulations. The idea of evolving CDM dynamics via this `Schr\"odinger-Vlasov correspondence' \citep{2018PhRvD..97h3519M} has been proposed since \cite{1993ApJ...416L..71W}, although
in practice simulations on classical computers commonly solve the Vlasov-Poisson equations directly via $N$-body methods \citep{2009MNRAS.398.1150B,2017ComAC...4....2P}. 

Our focus here is the Schr\"odinger-Poisson equations for quantum computer simulations because its form directly evolves a wavefunction, making it more more readily applicable to quantum simulation. 
We point out that the VQC method that we outline for solving the Schr\"odinger-Poisson equations in \S~\ref{sec:nonlinear} may generalized to other sets of nonlinear equations as well; the evolved quantity need not be a wavefunction but can be any smooth field.
An advantage of working with the Schr\"odinger-Poisson equation as opposed to the Vlasov-Poisson equation, however, is that the solution is smooth in the former, whereas the the Vlasov-Poisson equation often exhibits singularities, caustics, and discontinuities.


Initial conditions representing the early Universe are typically set by sampling a random Gaussian field with a prescribed power spectrum, and cosmological simulations may be evolved to the present day. 
A domain periodic in all spatial dimensions is commonly assumed.
Of particular interest is the evolution of the dark matter density 1D power spectrum, which is compared against observationally inferred values in order to constrain cosmological models \citep{2017PhRvD..96b3522I,2018arXiv180706209P}, 
as well as other global statistical properties such as the halo mass function or the subhalo abundance distribution \citep{2009MNRAS.398.1150B}.

\begin{figure}
\begin{center}
\includegraphics[width=0.47\textwidth]{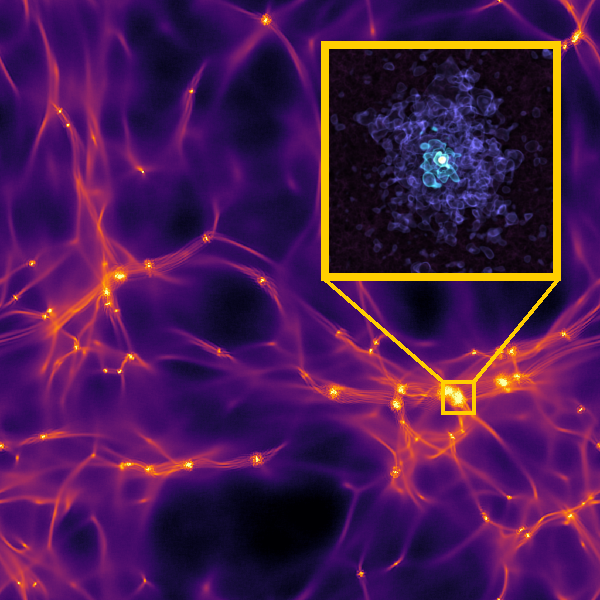}
\end{center}
\caption{A simulated cosmological volume from \protect\cite{2017MNRAS.471.4559M} evolved by solving the Schr\"odinger-Poisson equations using the classical algorithm of \S~\ref{sec:classical}. Shown is the log of projected density. The inset highlights an individual dark matter halo and its small-scale features.}
\label{fig:sim}
\end{figure}

\section{Classical algorithm}
\label{sec:classical}

The Schr\"odinger-Poisson equations (Eqns.~\eqref{eqn:S} and \eqref{eqn:P}) may be solved with a classical spectral method, which consists of applying unitary operators, Fast Fourier Transforms (FFTs), and inverse FFTs \citep{2017MNRAS.471.4559M}, and has been applied to cosmological simulations \citep{2017MNRAS.471.4559M,2019PhRvL.123n1301M,2020MNRAS.494.2027M}.

Algorithm \ref{algC} shows the steps of the spectral method. The system state $\bm{\psi}$ is an array of $N^3$ grid elements and $\bm{k}$ is an array of corresponding wavenumbers in the spectral decomposition. The system is evolved for $N_t$ timesteps. A `$.$' denotes an element-wise broadcast operator (e.g. used for multiplication or division).

\begin{algorithm}[H]
\caption{Spectral algorithm (classical)}\label{algC}
\begin{algorithmic}[1]
\For{$i = 1,\ldots,N_t$}
\State $\hat{\bm{\psi}} \gets \text{FFT}(\bm{\psi})$
\State $\hat{\bm{\psi}} \gets \exp(-i \frac{\bm{k}^2}{2} \Delta t).*\hat{\bm{\psi}}$
\State $\bm{\psi} \gets \text{iFFT}(\hat{\bm{\psi}})$
\State $\bm{\rho} \gets |\bm{\psi}|^2$
\State $\hat{\bm{V}} \gets -\text{FFT}(\bm{\rho}-1) ./ \bm{k}^2$
\State $\bm{V} \gets \text{iFFT}(\hat{\bm{V}})$
\State $\bm{\psi} \gets \exp(-i \bm{V}\Delta t).*\bm{\psi}$
\EndFor
\end{algorithmic}
\end{algorithm}

 Given a resolution of $N^3$ grid elements and $N_t$ timesteps, the method scales as $\mathcal{O}(N_t N^3 \log (N^3))$.  The method has second-order accuracy in time (the error is due to the non-commutative nature of the kinetic and potential operators in the Schr\"odinger-Poisson equation); the time-scale to which $\Delta t$ should be compared is $\hbar/(m |V_{\rm max}|)$ \citep{2018JCAP...01..009G}. Being a spectral method, it has exponential convergence properties in space (as long as the maximal wavenumber is resolved). Time evolution using the spectral method is unconditionally stable both for constant $V$ and for $V$ given by Eqn.~\eqref{eqn:P} \citep{2018JCAP...01..009G}, a constraint which is independent of the spatial resolution.  Thus $N_t$ need not scale with $N$.

Fig.~\ref{fig:sim} shows an example of a simulated cosmological volume; shown is projected density ($|\psi|^2$) for a simulation using $1024^3$ resolution elements \citep{2017MNRAS.471.4559M}.

\section{Quantum simulation}\label{sec:quantum}

\subsection{Basics: states and operators}\label{sec:basics}

We first introduce quantum states and operators and discuss the concept of `quantum advantage' over classical algorithms.  


A quantum computer with $n$ qubits encodes a quantum state $|\varphi\rangle$ that lives in an $N\equiv2^n$-dimensional complex Hilbert space: $|\varphi\rangle = \sum_{l=0}^{N-1}  c_l^{ } |l\rangle$,
where the $c_l$ are complex numbers satisfying the normalization condition $\sum_{l=0}^{N-1}|c_l^{ }|^2=1$. Each basis state $|l\rangle$ is specified by each of the $n$ qubits being in the state $|0\rangle$ or the state $|1\rangle$, i.e. the $2^n$ basis states range from $|0\rangle\otimes\ldots|0\rangle\otimes|0\rangle$
to $|1\rangle\otimes\ldots|1\rangle\otimes|1\rangle$ (commonly notated as $|0\ldots0\rangle$ to $|1\ldots1\rangle$).  

In the context of numerically solving the Schr\"odinger-Poisson equations, or any other differential equation, this allows for a much more efficient representation of a discretized function on $N=2^n$ points.  In the classical spectral method of \S~\ref{sec:classical}, the wavefunction at each timestep is represented as a length $N$ vector, but as a quantum state it can be represented using just $n=\log_2(N)$ qubits. E.g., for a discretized function on the interval $[0,1)$, the value of the function at location 
$x \equiv \text{binary}(l) = \frac{1}{2}\sum_{j=0}^{n-1}l_j^{ } 2^{-j}$,
with each $l_j^{ }$ in $\{0,1\}$, is the coefficient of the basis state $|l_{0},\cdots,l_{n-1}\rangle$.  

This more efficient representation of states can, for some problems, also lead to more efficient computations. Classically, even a simple operator like element-wise multiplication on a length-$N$ vector must take time at least linear in $N$, while the same operator acting on the quantum state could potentially take time only linear/polynomial in $n$, i.e., an exponential speedup.  However, some operators will still take time proportional to $N$, so it is essential to determine for a given problem whether an efficient implementation is indeed available.  

One way of measuring the complexity of implementing an operation on a quantum computer is the minimal number of 1- and 2-site fundamental gates into which the operation can be decomposed; the particular fundamental gates that can be used vary depending on the particular hardware implementation, but gates that are commonly available include the Hadamard gate
\begin{equation}
H \equiv 
\frac{|0\rangle+|1\rangle}{\sqrt{2}} \langle 0 | +
\frac{|0\rangle-|1\rangle}{\sqrt{2}} \langle 1 |,\label{eq:H}
\end{equation}
which implements a local change of basis on a qubit, phase-shift gates
\begin{equation}
R(\phi)\equiv 
|0\rangle\langle 0 | + 
e^{i\phi} |1\rangle\langle 1 |\label{eq:RZ}
\end{equation}
for some particular angles $\phi$, and the two-qubit controlled-NOT gate,
\begin{equation}
\text{CNOT} \equiv |00\rangle\langle 00|+|01\rangle\langle 01|+|10\rangle\langle 11|+|11\rangle\langle 10|,\label{eq:CNOT}
\end{equation}
which creates entanglement between qubits.  Two particular important phase shift gates are the S and T gates, with $\phi=\pi/2$ and $\pi/4$, respectively.  Indeed, the gate set $\{H,S,T,\text{CNOT}\}$ is universal, meaning that any gate can be arbitrarily well-approximated by some combination of these gates \citep{bravyi2005universal}.

A quantum circuit to implement some operation (e.g. one timestep in a differential equation) is built from some number of layers of these gates applied to the $n$ qubits encoding the state of the system; the number of layers is called the depth of the circuit.  Since multiple gates within a single layer can be applied in parallel, the time to run a circuit scales with the depth, making the depth a good measure of computational complexity; another useful measure is the total number of gates, which is larger by a factor of at most $n$.  

To demonstrate an exponential quantum speedup, we need not consider any particular fundamental gate set, since implementing a more general one-, two, or even three-qubit gate with a given accuracy will require a number of fundamental gates that does not scale with the system size.  Thus it will be sufficient for our purposes to demonstrate that there are circuits with polynomial depth in the number of qubits, regardless of the gates that appear in the circuit.  (If the hardware only allows gates between nearby qubits, an $m$-qubit gate may also require up to $\mathcal{O}(n^{m-1})$ swap gates to bring the qubits together, incurring an additional polynomial, but not an exponential, cost.)

\subsection{Quantum simulation of a linear Schr\"odinger equation}
\label{sec:linear}

Before considering the application of quantum simulation to the full Schr\"odinger-Poisson equations (Eqns.~\eqref{eqn:S} and \eqref{eqn:P}), we first discuss the simpler case of the one-dimensional Schr\"odinger equation (Eqn.~\eqref{eqn:S}) in a fixed potential $V$.  As this is a linear equation, the quantum simulation can be performed by directly applying a quantum circuit; we follow the algorithm laid out in \cite{2008AmJPh..76..657B}.
The method we describe here 
is similar to the classical algorithm and serves as a good introduction to quantum simulations.
However, unlike the classical analog, it 
cannot be easily extended to nonlinear equations, 
for which an entirely different algorithm 
will be needed (VQC; \S~\ref{sec:nonlinear}).
The inclusion of the discussion of the linear solver here helps highlight the unique challenges when it comes to designing quantum algorithms.

To represent the wavefunction $\psi(x,t)$ on the domain $[0,L]$ as a quantum state, we first discretize the spatial domain into $N=2^n$ equally spaced points $x_l\equiv l\Delta x$ for $l=0,\cdots,N-1$ (so $\Delta x \equiv L/N$), then set the value of $\psi$ at the point $x_l$ as the coefficient of basis state $|l\rangle$:
\begin{equation}
|\Psi\rangle = \frac{1}{\sqrt{N}} \sum_{l=0}^{N-1} \psi(x_l^{ },t) | l \rangle.\label{eq:Psi}
\end{equation}
The normalization factor $N^{-1/2}$ corrects for the fact that $\psi(x,t)$ is normalized to the system volume, $L$, while the quantum state is necessarily normalized to 1.

To perform one timestep, we follow essentially the same algorithm as in the pseudocode of \S~\ref{sec:classical}: Fourier transform, multiply by $\exp(-i k^2 \Delta t/2)$, inverse Fourier transform, and multiply by $\exp(-i V(x) \Delta t)$.  In other words, the algorithm must, for each timestep, compute
\begin{equation}
|\Psi\rangle \leftarrow
e^{-i V \Delta t}
\text{iQFT}\left(
e^{-i\frac{k^2}{2}\Delta t}\text{QFT}\left(|\Psi\rangle\right)\right)
\label{eqn:drift}
\end{equation}
Each of these operations (assuming a reasonable potential $V$) can indeed be efficiently implemented, as measured by the required circuit depth.

The `QFT' in Eqn.~\eqref{eqn:drift} is the well-known Quantum Fourier Transform (\citealt{2002quant.ph..1067C}), which can be performed efficiently \citep{hales2000improved} and maps:
\begin{equation}
\sum_{k=0}^{N-1}  f(k) |k\rangle \leftarrow
\sum_{l=0}^{N-1}  \hat{f}(l) |l\rangle 
\end{equation}
where 
\begin{equation}
\hat{f}(l) \equiv \frac{1}{\sqrt{N}} \sum_{k=0}^{N-1} e^{2\pi i k l/N}f(k)
\end{equation}
are the discrete Fourier Transform coefficients.
The number of gates in the QFT algorithm scales as $\mathcal{O}(n^2)$, in contrast with the $\mathcal{O}(2^nn)$ classical FFT.

The other basic operation required is multiplication by a function,
\begin{equation}
|x\rangle \leftarrow  e^{if(x)} | x\rangle.
\end{equation}
In the important case where $f(x)$ is a polynomial of degree $m$, this can be implemented with $n^m$ gates, each acting on $m$ qubits, as outlined in \cite{2008AmJPh..76..657B}.  Consider as an illustrative example the harmonic oscillator potential $V(x)=x^2$.  Write the discretized variable $x$ via binary expansion as $(\Delta x/2) \sum_{j=0}^{n-1} l_j 2^{-j}$,
where $k_j$ represents the $j$ qubit state.
Then $x^2 = (\Delta x/2)^2 \sum_{i,j=0}^{n-1} l_i l_j\, 2^{-(i+j)}$
and
\begin{equation}
e^{-iV(x) \Delta t} = \prod_{i,j=0}^{n-1}
\exp\left(-i \Delta t \left(\frac{\Delta x}{2}\right)^2 l_i l_j\, 2^{-(i+j)}\right).
\end{equation}
Each of the $n^2$ factors is implemented by a gate acting only on qubits $i$ and $j$.  Note that in combination with the QFT, the exact same method applies to the kinetic operator $f(k)\propto k^2$.  

The overall computation time for each timestep, assuming $e^{-iV(x)\Delta t}$ has an efficient implementation, is $\text{poly}(n)$, which is indeed an exponential speedup compared with the classical spectral method. As numerical stability does not require the number of timesteps to scale with spatial resolution, this exponential speedup also applies to the full computation of the final state from the initial one.  There remains the question of how to efficiently extract physical observables from the final state in, a discussion which we postpone to \S~\ref{sec:observables}.

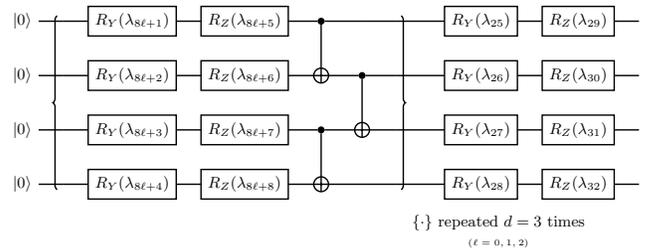
\begin{figure}
\begin{center}
\begin{adjustbox}{width=0.47\textwidth}
\begin{quantikz}
\lstick{$\ket{0}$} & \qw\lstick[wires=4]{} & \gate{R_Y(\lambda_{8\ell+1})} & \gate{R_Z(\lambda_{8\ell+5})} & \ctrl{1} & \qw      & \qw\rstick[wires=4]{\\ \\ \\ \\ \\ \\ \\ \\ \\ \\ \\ \\ \\ \\ $\left\{\cdot\right\}$ repeated $d=3$ times \\ {\tiny($\ell=0,1,2$)}} & \qw & \gate{R_Y(\lambda_{25})} & \gate{R_Z(\lambda_{29})} & \qw \\
\lstick{$\ket{0}$} & \qw & \gate{R_Y(\lambda_{8\ell+2})} & \gate{R_Z(\lambda_{8\ell+6})} & \targ{}  & \ctrl{1} & \qw & \qw & \gate{R_Y(\lambda_{26})} & \gate{R_Z(\lambda_{30})} & \qw  \\
\lstick{$\ket{0}$} & \qw & \gate{R_Y(\lambda_{8\ell+3})} & \gate{R_Z(\lambda_{8\ell+7})} & \ctrl{1} & \targ{}  & \qw & \qw & \gate{R_Y(\lambda_{27})} & \gate{R_Z(\lambda_{31})} & \qw  \\
\lstick{$\ket{0}$} & \qw & \gate{R_Y(\lambda_{8\ell+4})} & \gate{R_Z(\lambda_{8\ell+8})} & \targ{}  & \qw      & \qw & \qw & \gate{R_Y(\lambda_{28})} & \gate{R_Z(\lambda_{32})} & \qw \\
\end{quantikz}
\end{adjustbox}
\end{center} 
\caption{
Example of a quantum variational ansatz $U_j(\bm{\lambda})$ that uses $\mathcal{O}({\rm poly}(n))$ parameters $\bm{\lambda}$ to represent a solution field of size $2^n$.  Here $R_Y(\lambda)=\cos(\lambda)\text{Id}+i\sin(\lambda)\sigma_y$ and superposes $|0\rangle$ and $|1\rangle$; $R_Z(\lambda)=\cos(\lambda)\text{Id}+i\sin(\lambda)\sigma_z$ and is a phase rotation, equivalent to $R(-2\lambda)$ for $R$ defined in Eqn.~\eqref{eq:RZ}; and each two-qubit gate appearing in the third layer is CNOT defined in Eqn.~\eqref{eq:CNOT}.  The variational circuit has $(2d+2)n$ parameters, where $d$ is the circuit depth, and $d$ must be at least $\mathcal{O}(n)$ to represent a generic state.
}
\label{fig:VQC}
\end{figure}


\subsection{Cosmological (nonlinear) simulations with quantum-classical hybrid approach}
\label{sec:nonlinear}

We now consider a quantum algorithm for solving the full (nonlinear) Schr\"odinger-Poisson system.

It is not possible to straightforwardly adapt the linear quantum simulation algorithm of \S~\ref{sec:linear} to these nonlinear equations. 
One point of difficulty lies in the calculation of the density $|\psi|^2$, needed to obtain the self-potential. Unitary operators are unable to simply square coefficients of a quantum state, and the \textit{no-cloning} theorem means it is impossible to clone the state $|\Psi\rangle$.
Instead, we consider applying the hybrid quantum-classical VQC algorithm of \cite{2020PhRvA.101a0301L}.

In VQC, any state $|\varphi\rangle$ is given by a \textit{quantum circuit ansatz} $U(\bm{\lambda})$ that depends on some list of classical parameters $\bm{\lambda}$, $|\varphi\rangle\equiv|\varphi(\bm{\lambda})\rangle=U(\bm{\lambda})|0\rangle^{\otimes n}$; one commonly used hardware-efficient ansatz \citep{o2016scalable,shen2017quantum,2017Natur.549..242K} for the circuit $U$ is shown in Fig.~\ref{fig:VQC}.  It is easy to obtain multiple copies of any given ansatz state: the parameters $\bm{\lambda}$ are classical and can simply be stored and then used as many times as needed to generate more copies.  It is also easy to generate the complex conjugate $|\varphi^\ast\rangle$ when the parameters for $|\varphi\rangle$ are known---for example, in the ansatz from Fig.~\ref{fig:VQC}, one simply multiplies the classical parameters for all $R_Z$ gates by -1.

To simulate the Schr\"odinger-Poisson system, the wavefunction $\psi$ and potential $V$ at each timestep are each given by just such an ansatz state, $|\Psi(\bm\lambda)\rangle$ and $|V(\bm{\tilde\lambda})\rangle$ respectively.  The computational task in each timestep is to first find a new set of parameters $\bm{\tilde\lambda}$ that give the best approximation to the self-potential due to the current wavefunction, then to find a new set of $\bm\lambda$ that give the best approximation to the new state after the timestep.  Each of these optimizations is performed classically, for example using standard approaches such as Nelder-Mead, quasi-Newton methods or others, \citep{spall1998overview}, with the cost function computed using quantum circuits; it is important to note that each cost function involves a measurement of the quantum state, and thus for each set of classical parameters it must be evaluated a number of times.  Also note that because the potential is real-valued, we may remove all $R_Z$ gates from the variational circuit for $|V\rangle$.  The algorithm is summarized in Algorithm~\ref{algQ2}. 


\begin{algorithm}[H]
\caption{Quantum VQC Algorithm}\label{algQ2}
\begin{algorithmic}[1]
\For{$i = 1,\ldots,N_t$}
\While{$\{\bm{\tilde\lambda}_i,\tilde{\lambda}_V\}$ not converged}
\State $|\Psi\rangle,\,|\Psi^\ast\rangle \gets \bm{\lambda}_{i-1};\,\,|V\rangle \gets \bm{\tilde\lambda}_i$
\State $\text{Cost} = ||\,(\bm{|\psi|^2}-\bm{1}) - \tilde{\lambda}_V\nabla^2 \bm{V}\, ||^2$
\EndWhile
\While{$\bm{\lambda}_i$ not converged}
\State $|\Psi\rangle \gets \bm{\lambda}_{i-1};\,\,|V\rangle \gets \bm{\tilde\lambda}_i$;\,\,$|\psi^\prime\rangle \gets \bm{\lambda}_i$
\State $\text{Cost} = ||\,(\bm{\psi}^\prime - \bm{\psi}) + i\Delta t(\tilde{\lambda}_V\bm{V}+\nabla^2)\bm{\psi}^\prime\,||^2$
\EndWhile
\EndFor
\end{algorithmic}
\end{algorithm}

At each timestep we first find the variational parameters $\bm{\tilde\lambda}$ for the state representing the potential, $|V\rangle$.  We also need an additional variational parameter $\tilde{\lambda}_V$ that normalizes $|V\rangle$ to the physical value of the potential.  
(We present the cost functions in terms of the vectors represented by the quantum states, rather than the states themselves, but will clarify below exactly how the cost functions can be computed from the states in practice.)  If the ansatz for $|V\rangle$ is fully general, then the cost function will be minimized at 0 when the Poisson equation (Eqn.~\ref{eqn:P}) is solved exactly.
Once the new parameters for the potential have been determined, we complete the timestep by optimizing the new parameters for $|\Psi\rangle$, using the just-computed potential in the cost function.

What remains to be discussed is the computation of cost functions, lines 4 and 7 of Algorithm~\ref{algQ2}.  For each cost function, we break it up into many terms, each of which is a vector product, and then for each term design a quantum circuit whose average output over many runs will give the value of that vector product (with an additional factor that can be included classically at the end).

The first step is to replace the spatial derivatives $\nabla^2$ of $\bm{V}$ and $\bm{\psi}^\prime$ by discretized derivatives.  In one dimension, $\nabla^2\bm{\psi}^\prime$ for example becomes $\left(\bm{\psi}_{-1}^\prime-2\bm{\psi}^\prime_{ }+\bm{\psi}^\prime_{+1}\right)/\left(\Delta x\right)^2$, where $\bm{\psi}^\prime_{\pm 1}=\bm{\psi}^\prime_{ }(x\pm\Delta x)$; the quantum state for $\bm{\psi}^\prime_{\pm 1}$ is $|\Psi^\prime_{\pm 1}\rangle = {\frac{1}{\sqrt{N}}\sum_l\psi^\prime_{ }(x_l)|l\mp 1\rangle}$.
An explicit shift circuit to compute, e.g., $|\Psi_{+1}\rangle$ from $|\Psi\rangle$ by shifting each coefficient to the previous basis state is given by Fig.~S2 of \cite{2020PhRvA.101a0301L} with the unseen control qubit set to $|1\rangle$.  
The resulting cost function for the potential becomes ${||\,(\bm{|\psi|^2}-\bm{1}) - (\tilde{\lambda}_V/\Delta x^2)(\bm{V}_{-1}^{ }-2\bm{V}+\bm{V}^{ }_{+1})\, ||^2}$, with the cost function for $\bm{\psi}^\prime_{ }$ likewise modified.

We can now divide the terms appearing in the two cost functions into four types: the four-state product $|\bm{\psi}|^2\cdot|\bm{\psi}|^2$; the three-state products $|\bm{\psi}|^2\cdot\bm{V}$ and $|\bm{\psi}|^2\cdot\bm{V}_{\pm 1}^{ }$; the single-vector sums $\bm{1}\cdot\bm{V}$ and $\bm{1}\cdot\bm{V}^{ }_{\pm 1}$; and many two-state products, e.g. $\bm{\psi}^\ast_{ }\cdot\bm{\psi}^\prime_{ }$.

A circuit for $|\bm{\psi}|^2\cdot|\bm{\psi}|^2$ is given explicitly in \cite{2020PhRvA.101a0301L}, Figs.~1(a) and 2(a).  Likewise the three-state products can be computed using the circuits in Figs.~1(a) and 2(b) of the same reference, with the box labeled ``V'' in 2(b) being the variational circuit for $|V\rangle$ followed by, for the $\bm{V}_{\pm 1}^{ }$ terms, the appropriate shift circuit as discussed above.  Note that the results of these circuits must also be multiplied by scalar factors, for example $N^2$ for $|\bm{\psi}|^2\cdot|\bm{\psi}|^2$, to get the final cost function contributions.

Two-vector products like $\bm{\psi}^\ast_{ }\cdot\bm{\psi}^\prime_{ }$ are found using the state overlap $\langle \Psi|\Psi^\prime_{ }\rangle$.  However, some care is required because algorithms like the well-known swap test circuit~\citep{buhrman2001quantum} for state overlap return the magnitude of the overlap, but in our case we need $\langle \Psi|\Psi^\prime_{ }\rangle + \langle \Psi^\prime_{ }|\Psi\rangle$ which in general is different from $|\langle \Psi|\Psi^\prime_{ }\rangle| + |\langle \Psi^\prime_{ }|\Psi\rangle|$.  Fortunately, since $\langle \Psi|\Psi^\prime_{ }\rangle + \langle \Psi^\prime_{ }|\Psi\rangle$ is real, it is sufficient to find the real part of $\langle \Psi|\Psi^\prime_{ }\rangle$.  This can be computed using the Hadamard test circuit~\citep{nielsen2002quantum}, which for a state $|\varphi\rangle$ and unitary operator $U$ returns the real part of $\langle \varphi | U | \varphi\rangle$.  In our case, for $|\varphi\rangle$ we use the $n$-qubit state $|\bm{0}\rangle$ and for the unitary we use $U(\bm{\lambda})^\dg_{ }U(\bm{\lambda}^\prime_{ })$, where $U(\bm{\lambda})$ is the variational circuit for $|\Psi\rangle$ and $U(\bm{\lambda}^\prime_{ })$ for $|\Psi^\prime_{ }\rangle$; note that $U^\dg_{ }$ can be found by running the circuit for $U$ in reverse, taking the Hermitian conjugate of each $R_Z$ and $R_Y$ gate.  This same circuit is also given in \cite{2020PhRvA.101a0301L}, Figs.~1(a) and S3(a).  With this approach, all terms in the cost function for the wavefunction can be organized into complex conjugate pairs, and the total contribution of the pair can be computed by finding the real part of one of the two terms; again, a scalar multiple is also needed for each term.

Finally, the single-vector sums can also be found by the overlap of two states.  In this case, the vector $\mathbf{1}$ can be represented by the quantum state $\left(H|0\rangle\right)^{\otimes n}$, which has all coefficients equal to $2^{-n/2}$, along with a scalar multiple of $\sqrt{N}$ as a correction.  Here, since $\bm{V}$ is real, finding the magnitude of the overlap for each term is sufficient, so the swap test can be used rather than the more general Hadamard test approach outlined above.

We thus can construct an explicit circuit for each term in the cost functions, and the value of the term is given by the average of a measurement made after each of many runs of the circuit.  For $M$ circuit evaluations, the error scales as $M^{-1/2}$; in the large-$n$ limit, the relative error also grows linearly in $N$, which would seem to require an exponentially large number of samples.  Fortunately, this scaling only holds once $n$ is large enough to converge all spatial features, so in the regime of physical interest it does not present a problem \citep{2020PhRvA.101a0301L}. 

This variational quantum algorithm gives an exponential speedup over classical methods.  This is true for each timestep, since the cost functions are evaluated by circuits with depth polynomial in $n$, and,  as just noted above, the number of circuit evaluations $M$ needed for converged measurement statistics for the cost functions only scales with  $n$ once $n$ is large enough to resolve all features of the solution~\citep{2020PhRvA.101a0301L}.  Furthermore, with a good choice of optimization algorithm, it may not be necessary to converge the measurement statistics for each set of parameters~\citep{Sweke2020stochasticgradient, Kubler2020adaptiveoptimizer}.  The last consideration for the cost of each individual timestep is the number of iterations needed to converge the classical optimization of the parameters.  While some variational quantum algorithms suffer from the barren plateau problem~\citep{cerezo2020variational}, leading to an exponential number of iterations, this should not occur when the optimization is initialized close to the solution, which is the case in each timestep because we start with the parameters from the previous step and time evolution is continuous.  Consequently, we expect that each timestep has total cost poly($n$).  Finally, for time-stepping we use a first-order implicit Euler method, so that numerical stability does not require the number of timesteps to scale with spatial resolution.  As the actual time evolution produced by the simulation is the same as in the classical case, that the number of timesteps also need not scale with spatial resolution to achieve a numerically accurate solution carries over to the quantum simulation.  As a result, the full simulation is also exponentially faster than for classical methods. 

To conclude this section, we offer some important notes: (1) We use a second-order discretization for the Laplacian operator, and a first-order implicit Euler method for timestepping, but higher-order schemes are possible.  
 (2) More efficient circuits for certain cost functions are possible.  There has been some recent progress in using machine learning to optimize quantum circuits, including for computing overlaps \citep{Cincio_2018}.
 (3) As noted above, when optimizing the cost functions, in practice it may not be necessary to converge measurement statistics for each value of $\bm{\lambda}$ or $\bm{\tilde\lambda}$.  As demonstrated in \cite{Sweke2020stochasticgradient} and \cite{Kubler2020adaptiveoptimizer}, instead one can make only a small number of measurements for each term and use stochastic gradient descent, which provides a substantial speedup.

\subsection{Efficient computation of observables}\label{sec:observables}

One major challenge in quantum computing is extracting useful information at the end of a simulation.  In a classical calculation, the full state with $N=2^n$ coefficients is known and can be used for further computations, for example of expectation values and other physical observables.  The output of a quantum circuit, however, is much less accessible.  Measuring each qubit will return one of the $2^n$ basis states, and running the circuit and measuring its output many times will then give an approximate probability distribution over the basis states, in other words an approximation to $|\psi|^2$.  This is, however, not practical.  If any extensive fraction of the basis states have a non-negligible weight, it will require exponentially many measurements to extract a reasonable approximation, entirely defeating the purpose of using a quantum simulation in the first place.  One could sample a much coarser grid, so that measurement statistics converge with a reasonable number of samples; this, however, loses important information contained in the final wavefunction.  Alternatively, in the case of the variational algorithm, the exact state could in principle be reconstructed classically from the parameters at any given timestep.  While using this approach to find the full final state after completing the variational quantum simulation would still be faster than performing a classical simulation, and depending on the output state could be more efficient than sampling many copies of the quantum state, it still incurs an exponential cost in $n$ and hence is not practical if $n$ is large enough that a quantum simulation is useful in the first place.

A better approach is to directly consider the quantities of physical interest extractable from the wavefunction, and to use quantum circuits that, with the final wavefunction as input, end with a measurement on just one or a few qubits that will give the desired quantity.  Specializing to the dark matter simulations we discuss here, one particularly important quantity that can be compared with observational data is the power spectrum, i.e. the Fourier transform of $\psi$ as a function only of the radial component of $\bm{k}$.  While this still consists of $\mathcal{O}(2^n)$ numbers and hence is not efficiently measurable, moments of the power spectrum are easily accessible.

As discussed in \S~\ref{sec:nonlinear}, there are efficient quantum circuits to compute products of wavefunctions, and these can be used to compute expectation values.  The other key is that there are efficient quantum circuits to prepare states representing any polynomial: a polynomial of degree $m$ can be represented exactly as a matrix product state with bond dimension $m+1$ \citep{Oseledets2013}, which can be converted into a low-depth quantum circuit \citep{2020PhRvA.101a0301L}  By using the product circuit with the state representing $x^m$, one copy of $|\Psi\rangle$, and one copy of $|\Psi^\ast\rangle$, the resulting measurements average to $\langle x^m \rangle$.  This allows efficient computation of any moment $\langle x^l y^m z^p\rangle$.  Moments of the power spectrum can be similarly obtained by first applying the QFT to both $|\Psi\rangle$ and $|\Psi^\ast\rangle$, in which case one would instead get $\langle k_x^l k_y^m k_z^p\rangle$.  Other observables may be efficiently computable using techniques from machine learning \citep{kiani2020quantum}.

Computation of observables in this manner is substantially faster using a variational method as in \S~\ref{sec:nonlinear} rather than a direct simulation as in \S~\ref{sec:linear}. In the former case, one need only run one copy of the low-depth variational circuit for the state in order to measure it, while in the latter case one must run the full simulation each time.  This is one reason to consider using a variational method even if direct simulation is possible.  Another, related benefit to using the variational approach is that the variational parameters can be stored classically not just for the final state, but indeed at each timestep, with relatively little memory, namely $N_t\,{\rm poly}(n)$.  Then observables can later be efficiently calculated for the state at any intermediate timestep, just as for the final state.


\section{Mock quantum simulation}
\label{sec:mock}

We perform a mock simulation of the quantum Algorithm~\ref{algQ2} (\S~\ref{sec:nonlinear}) where we simulate the entire quantum state on a classical computer and apply the quantum circuit in each step as a matrix multiplication. We consider the 1D Schr\"odinger-Poisson system on the domain $[0,8]$ with initial condition $\psi(x,0) = \sqrt{1 + 0.6\sin(\pi x/4)}$ and evolve it until $t=3$. In this simple problem, the overdensity in the initial condition collapses due to self-gravity and then splits into two during this time interval.

\begin{figure}
\begin{center}
\includegraphics[width=0.47\textwidth]{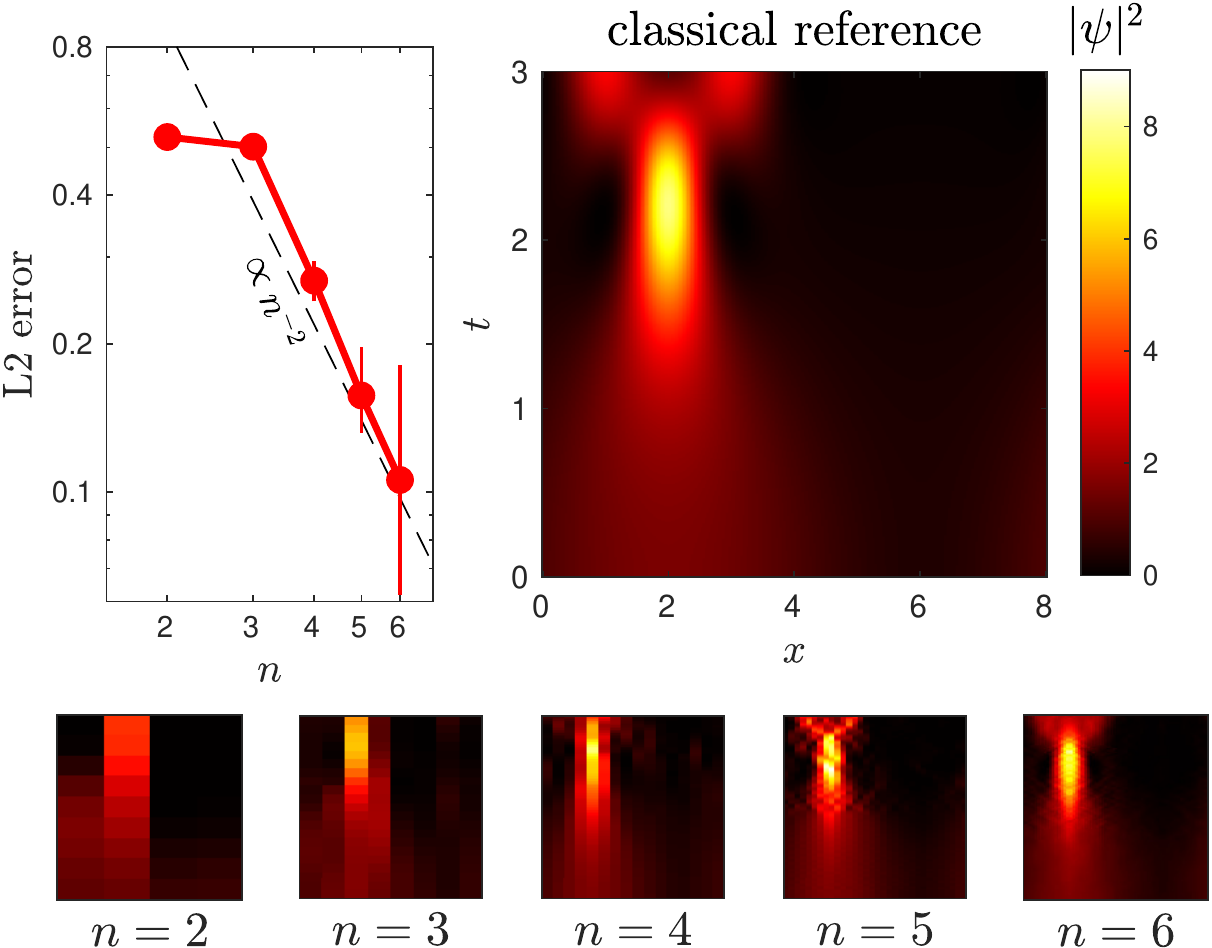}
\end{center}
\caption{Proof-of-concept mock simulations of the quantum algorithm for the nonlinear 1D Schr\"odinger-Poisson system with $n=2,3,4,5,6$ qubits and a reference solution used to demonstrate numerical convergence.  Error bars show one standard deviation for 20 samples with $\mathcal{O}(10^{-5})$ perturbations to variational parameters at each timestep.}
\label{fig:mockSim}
\end{figure}

Fig.~\ref{fig:mockSim} shows the results of the simulation for $n=2,3,4,5,6$ qubits, and an error analysis of numerical convergence to the classical reference solution carried out on an $N=256$ grid.  Note that we do not simulate the effect of measurement statistics, either in computing cost functions or in representing the final state; rather, we make use of our classical access to the full wavefunction. 

 A quasi-Newton (derivative-free) algorithm was used for minimization, and the quantum variational ansatz was taken to be of form Fig.~\ref{fig:VQC} with depth $d=n-1$. We have not optimized the choice of minimization algorithm or the form of the variational ansatz for our proof-of-concept demonstration, although our choice works reasonably well.
 
 Different types of noise affect a quantum calculation, for which noise-resilient techniques are being developed \citep{fontana2020evaluating}. We do not consider full hardware and measurement noise, but investigate the effect of adding random Gaussian errors to the variational parameters ($\lambda\leftarrow \lambda+\delta \lambda$) after each timestep to examine how errors may propagate during the nonlinear time evolution. Our simulations are robust to perturbations with  $\delta\lambda\lesssim\mathcal{O}(10^{-5})$ (Fig.~\ref{fig:mockSim}).

\section{Future prospects and concluding remarks}
\label{sec:conc}

Emerging quantum computing technology has the potential to revolutionize cosmological simulations.
Historically, simulation size has grown exponentially with time following Moore's Law (Fig.~1 of \citealt{2014MNRAS.445..175G}). Quantum hardware may allow for even faster growth, or at least avoid exponentially growing energy costs.

We have outlined a VQC algorithm for efficiently evolving the \textit{nonlinear} Schr\"odinger-Poisson equations for the cosmological simulation of dark matter, as well as a quantum algorithm for simulating \textit{linear} Schr\"odinger systems which also has applications to certain problems in astrophysics \citep{2020JCAP...01..001L}.  
Both algorithms ($\mathcal{O}(N_t {\rm poly} (n))$) offer an exponential speedup over classical simulations ($\mathcal{O}(N_t 2^nn)$). Physically relevant observables can be efficiently computed in the quantum simulation and compared against astrophysical observations.  The VQC method may be generalizable to other sets of differential equations as well.

Our proof-of-concept mock simulation for nonlinear problems demonstrates that the quantum algorithm works in principle.  In practice, current technology is limited by high error rates in two-qubit gates and fast loss of coherence due to interactions with the environment, making the algorithms inaccessible for the time being.  However, with limited-depth circuits, our algorithm may become feasible even before the advent of fully error-corrected quantum computation. 

Further improvements are certainly possible, including optimized choices for the quantum variational ansatz and minimization algorithm, as well as more efficient circuits for computing the cost functions.  A more exotic possibility is that of a direct, rather than variational, quantum algorithm for the simulation of the Schr\"odinger-Poisson equations.
The equations preserve information and are reversible, so if any nonlinear system can be directly simulated, it is a strong candidate.

\section*{Acknowledgments}
We thank Juan Miguel Arrazola and Timothy Hsieh for helpful comments on earlier versions of this manuscript, as well as the anonymous referee's comments which helped improve our discussion on certain aspects of the quantum circuits used in our VQC algorithm. Support (PM) for this work was provided by NASA through Einstein Postdoctoral Fellowship grant number PF7-180164 awarded by the \textit{Chandra} X-ray Center, which is operated by the Smithsonian Astrophysical Observatory for NASA under contract NAS8-03060.
Research at Perimeter Institute is supported in part by the Government of Canada through the Department of Innovation, Science and Economic Development Canada and by the Province of Ontario through the Ministry of Economic Development, Job Creation and Trade.

\bibliography{mybib}{}

\end{document}